\documentclass[9pt,conference]{IEEEtran}
\IEEEoverridecommandlockouts

\usepackage{cite}
\usepackage{amsmath,amssymb,amsfonts}
\usepackage{algorithmic}
\usepackage{graphicx}
\usepackage{textcomp}
\usepackage{xcolor}

\usepackage{subcaption}
\usepackage{multirow}
\usepackage{booktabs}
\usepackage{url}

\def\BibTeX{{\rm B\kern-.05em{\sc i\kern-.025em b}\kern-.08em
    T\kern-.1667em\lower.7ex\hbox{E}\kern-.125emX}}
\begin{document}

\title{XLSR-Transducer: Streaming ASR for Self-Supervised Pretrained Models}

\author{
    \parbox{\linewidth}{%
        \centering Shashi Kumar$^{1,2}$, 
        Srikanth Madikeri$^{1,3}$, Juan Zuluaga-Gomez$^{1,2}$,  
        Esaú Villatoro-Tello$^{1}$,\\
        \centering Iuliia Thorbecke$^{1,3}$, Petr Motlicek$^{1,4}$, Manjunath K E$^{5}$ and Aravind Ganapathiraju$^{5}$ \\[1ex]
        \small $^{1}$ \textit{Idiap Research Institute, Martigny, Switzerland} \\
        \small $^{2}$ \textit{EPFL, Lausanne, Switzerland} \\
        \small $^{3}$ \textit{University of Zurich, Switzerland} \\
        \small $^{4}$ \textit{Brno University of Technology, Czech Republic} \\
        \small $^{5}$ \textit{Uniphore, India} \\
    }
}


\maketitle

\begin{abstract}
Self-supervised pretrained models exhibit competitive performance in automatic speech recognition on finetuning, even with limited in-domain supervised data.
However, popular pretrained models are not suitable for streaming ASR because they are trained with full attention context.
In this paper, we introduce XLSR-Transducer, where the XLSR-53 model is used as encoder in transducer setup.
Our experiments on the AMI dataset reveal that the XLSR-Transducer achieves 4\% absolute WER improvement over Whisper large-v2 and 8\% over a Zipformer transducer model trained from scratch. To enable streaming capabilities, we investigate different attention masking patterns in the self-attention computation of transformer layers within the XLSR-53 model.
We validate XLSR-Transducer on AMI and 5 languages from CommonVoice under low-resource scenarios.
Finally, with the introduction of attention sinks, we reduce the left context by half while achieving a relative 12\% improvement in WER.
\end{abstract}


\section{Introduction}
\begin{figure}[t]
    \centering
    \includegraphics[width=0.99\linewidth]{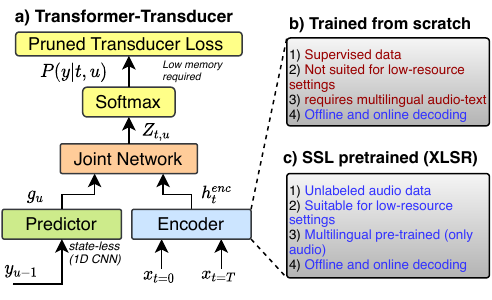}
    \caption{Current state-of-the-art a) Transducer ASR includes state-less predictor, pruned transducer loss and b) Transformer-based encoder, trained from scratch. We replace the encoder by XLSR-53, an SSL model suitable for low-resource applications. Our contributions lead to the c) XLSR-Transducer.}
    \label{fig:xlsr-transducer}
    \vspace{-0.5cm}
\end{figure}

End-to-End (E2E) modeling approaches for automatic speech recognition (ASR) have become ubiquitous in literature.
There are three popular architecture for achieving E2E models: encoder-decoder (AED) based models \cite{whisper, watanabe2017hybrid, mai23_interspeech,prabhavalkar2023end}, using Connectionist Temporal Classification (CTC) loss \cite{ctc_loss}, and neural Transducer \cite{graves2012sequence,pruned-rnnt-loss} based models.
The transducer models, shown in Figure~\ref{fig:xlsr-transducer}a, consist of an encoder, predictor and joint networks.
Using Transformers \cite{vaswani2017attention} encoder, termed as Transformer Transducer (TT)~\cite{yeh2019_tt,zhang2020_tt}, is a popular choice for streaming ASR \cite{cascaded-encoder-streaming,fastconformer-streaming,rnnt-las-rescoring} because of its inherently streaming nature. 
It is common to train TT models from scratch which requires large amount of supervised data \cite{fastconformer-streaming,cascaded-encoder-streaming}.
Self-supervised (SSL) models have shown strong ASR performance when trained with small amount of in-domain supervised data \cite{xlsr,mms}.
Most of the recent works using SSL models for ASR use encoder-decoder setup \cite{wav2vec2-aed, w2v-bert-aed, speecht5-aed} or CTC loss \cite{xlsr,mms,wav2vec2-ctc-multilingual, mslam-pretraining-ctc}.
In this paper, we present XLSR-Transducer where we use a pretrained XLSR-53 model as encoder in the TT architecture. This opens the door to streaming TT systems for low-resource applications.

In streaming ASR, partial hypotheses are generated for each audio chunk sequentially \cite{fastconformer-streaming,swietojanski2023variable_attention} to produce the transcript for the full audio, whereas the entire audio segment is available for non-streaming decoding.
Depending on the latency requirements, the chunk size may vary from few hundred milliseconds to few seconds \cite{swietojanski2023variable_attention}.
Typically, a drastic degradation in word error rate (WER) is observed when non-streaming models are decoded in streaming fashion \cite{rnnt-las-rescoring}, because only a limited context is available.
In this work, we propose a variety of attention masking patters that enables streaming training and decoding of our XLSR-Transducer model.
We also study the importance of chunk sizes and past context by varying them during inference.
Increasing past context typically enhances ASR performance \cite{swietojanski2023variable_attention}, at the expense of increased latency.
Recently, it was shown that the transformer layers learn to assign disproportionate attention scores to few initial tokens for streaming language models \cite{attn-sink-main}, termed as attention sinks.
We study the effects of attention sinks for the first time in streaming ASR.
Formally, at decoding time, we allow the transformer layers in XLSR to attend to few initial frames in addition to designated frames in chunk and past context. In theory, this reduces total computation required for processing an audio chunk during streaming decode.
Although SSL encoder-based transducer models have been explored in \cite{best-rq}, detailed results for varying chunk sizes and past context are not available. The trade-off between increased latency due to larger chunk or past context sizes and performance is not immediately clear, especially when the SSL model is not pre-trained in a streaming fashion. In this work, we present a comprehensive analysis that can better inform the choice of these hyperparameters based on specific requirements. Moreover, our work explores attention sinks, which can reduce overall computation while improving performance.

\begin{figure*}[t]
    \centering
    \includegraphics[width=0.99\linewidth]{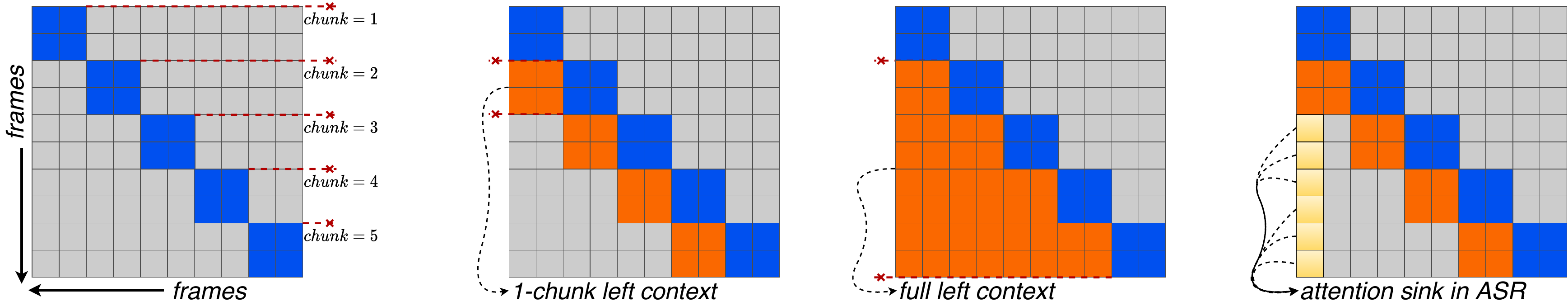}
    \caption{Masking strategies for streaming XLSR-Transducer. Multi-chunk training allows decoding with variable chunk size (blue) and left context (orange). Each square denotes “n” frames. Attention sink (yellow) allows context from the first n frames. Our results show that \textbf{attention sink frames offer a better trade-off w.r.t increasing left context alone}, leading to lower WERs.
    }
    \label{fig:xlsr-transducer-mask}
    \vspace{-0.3cm}
\end{figure*}

\noindent \textbf{Our contributions are covered below:}

\begin{itemize}
    \item Introduction of the XLSR-Transducer, a multilingual SSL encoder based transducer model, demonstrating significant WER improvement on the AMI dataset compared to large speech foundational models and other open-source ASR models;
    \item extension to streaming XLSR-Transducer and a systematic study of chunk size and past context on training and inference;
    \item to author's knowledge, this is the first work that explores the attention sink~\cite{attn-sink-main} phenomenon for streaming ASR which leads to improved WER; and
    \item Evaluation of the XLSR-Transducer on AMI and five languages of CommonVoice dataset in low resource settings.
\end{itemize}

\section{XLSR-Transducer}
\label{sec:xlsr-transducer}

In a typical Transformer-Transducer (TT) ASR model (Figure \ref{fig:xlsr-transducer}a), there are three networks: the encoder, predictor, and joiner. The encoder processes audio frames to produce acoustic embeddings.
The predictor generates token embeddings in an auto-regressive manner, taking previous non-blank tokens as input. Lastly, the joiner combines the outputs from the encoder and predictor to predict a probability distribution over the tokens in the vocabulary.
In this work, we utilize a stateless predictor \cite{stateless-predictor} composed of an embedding layer and one 1-D CNN layer, and the joiner network consists of one linear layer. 
Typically, the encoders \cite{lstm,conformer,zipformer,fastconformer} are trained from scratch and require a large amount of in-domain supervised data to achieve decent WER, which may not always be feasible. 
We train the TT model using the pruned-transducer loss \cite{pruned-rnnt-loss} from k2\footnote{\url{https://github.com/k2-fsa/k2}} toolkit. 

\subsection{Non-Streaming XLSR-Transducer}
\label{subsec:xlsr-transducer}
In contrast to encoders trained from scratch, recent advancements in SSL pretrained models demonstrate competitive performance \cite{mms,xlsr} when fine-tuned with a limited amount of labeled data for ASR. 
Previously, the popular ASR systems employing SSL pretrained models have utilized CTC loss~\cite{ctc_loss}, encoder-decoder based architecture~\cite{watanabe2017hybrid,mai23_interspeech}, and Lattice-Free MMI loss \cite{lfmmi-wav2vec2} (hybrid approach) for training.
In this paper, we integrate pretrained models as encoders in the TT setup, as illustrated in Figure \ref{fig:xlsr-transducer}c.
One notable advantage is the ability to achieve strong ASR performance with relatively low amounts of training data.
We select XLSR-53 \cite{xlsr} as our encoder model, which takes raw audio as input and outputs audio frames with a frame duration of 25ms and a stride of 20ms.
The selection of XLSR-53 is driven by its large-scale pre-training on multilingual audio, which has demonstrated competitive ASR performance \cite{xlsr} in the low-resource across multiple languages.

\subsection{Streaming XLSR-Transducer}
\label{subsec:streaming-xlsr-transducer}

XLSR is typically trained and decoded using entire audio sample. 
This makes the proposed XLSR-Transducer non-streaming despite the use of stateless predictor and linear joiner that are inherently streaming.
The main challenge to port SSL models to streaming is the use of self-attention in the transformer layers~\cite{vaswani2017attention}, i.e., computed over entire acoustic frames of an utterance. In this paper, we present multiple masking patterns~\cite{swietojanski2023variable_attention} to limit the frame context over which self-attention is computed, simulating streaming ASR within the XLSR-Transducer setup.

\noindent \textbf{Chunked masking} \quad In streaming ASR, decoding partial hypotheses begins after receiving a few audio frames, known as the chunk size. As depicted in \mbox{Figure \ref{fig:xlsr-transducer-mask}a}, we implement chunk-wise decoding by masking frames outside a specific chunk during the forward pass from the XLSR model.
The mask is applied after dot-product computation during self-attention, ensuring that each frame inside a chunk has access to all the frames within that chunk.
Note that the XLSR model also includes a CNN front-end, which takes raw audio as input. Thus, we feed chunk-size equivalent raw waves to the CNN front-end in a sequential manner and concatenate them across the time dimension to obtain all the frames for an utterance.
In this paper, we explore chunk sizes of 16, 32, 64, and 128, translating to approximately 320ms, 640ms, 1280ms, and 2560ms, respectively, for XLSR.

\noindent \textbf{Chunked masking with variable left context chunks} \quad In practice, when decoding chunk ``n", we have access to all the previous chunks, which can be utilized as left context.
As illustrated in Figures \ref{fig:xlsr-transducer-mask}b and \ref{fig:xlsr-transducer-mask}c, a variable number of left context chunks can be utilized during the self-attention computation of a chunk, with the possibility of using the full left context.
The number of frames in the left context is a multiple of the chunk size, as this can be efficiently implemented to store past chunks in the cache.

\noindent \textbf{Streaming training and decoding}  \quad The use of non-streaming XLSR-Transducer for streaming decoding with the described masking patterns presents a challenge. The model has been trained on full context, creating a train-test mismatch.
To address this challenge, we train the model in a streaming fashion using a fixed chunk size and left context.
Flexibility in our chunked mask implementation allow us to perform both streaming and non-streaming decoding using a single model.
The advantage of our method is that it only affects the fine-tuning stage and we can avoid the computationally prohibitive pre-training of the SSL model.

\noindent \textbf{Multi-chunk Training} \quad In many practical use-cases of streaming ASR, varying the chunk size at inference is often desirable depending upon latency requirements. However, a streaming XLSR-Transducer model trained with a fixed chunk size may not yield optimal WERs when decoded with different chunk sizes. Also, training multiple models for varying chunk sizes may be infeasible.
To address this limitation, we propose randomly selecting the chunk size from the predefined list mentioned above for each batch during training.

\subsection{Attention sinks for streaming ASR}
In a recent work on streaming language models (LM) \cite{attn-sink-main,big-bird-attn-sink,lm-infinite}, it was shown that a surprisingly large amount of attention scores during self-attention computation inside transformer layers is directed towards the initial tokens, termed as \textit{attention sinks}. 
This was attributed to the Softmax operation, which mandates attention scores to sum up to one and in autoregressive LMs, all subsequent tokens have access to the initial tokens.
Consequently, the model may find it easier to learn to assign large scores to these initial tokens.
In our streaming model training, where we utilize full left context, we employ a similar setup.
This leads us to introduce the first utilization of the attention sinks in the context of streaming ASR during inference.
Specifically, as depicted in Figure \ref{fig:xlsr-transducer-mask}d, we enable self-attention to focus on not only frames within a chunk and left context chunks, but also on the initial few frames.

\section{Experimental setup}
\label{sec:exp}

\subsection{Datasets}
\label{subsec:datasets}
We benchmark our proposed model on two datasets—AMI and CommonVoice.
We choose AMI, instead of Librispeech \cite{panayotov2015librispeech}, because it involves natural, multi-party, conversational speech with acoustic complexity and overlapping dialogue, whereas LibriSpeech contains read speech, making AMI more relevant for real-world scenarios.

\noindent \textbf{AMI} \quad We train and evaluate the XLSR-Transducer model on the individual head microphone (IHM) split from the AMI dataset \cite{ami} containing audios with a sampling rate of 16\,kHz.
We use the default recipe for AMI from lhotse\footnote{\url{https://github.com/lhotse-speech/lhotse}} toolkit to prepare the train, dev and eval sets containing 80hr, 8.8hr, 8.5hr of audios respectively.
In all our experiments on the AMI dataset, we use WER on the dev set to select the best epoch and report the results on the eval set.

\noindent \textbf{CommonVoice} \quad We validate XLSR-Transducer on five non-English languages from CommonVoice-v11~\cite{ardila2020common}.\footnote{CommonVoice-v11: cv-corpus-11.0-2022-09-21.} This includes Catalan (CA), Belarusian (BE), Spanish (ES), French (FR) and Italian (IT). To keep experimentation under the low-resource domain, we extract randomly a 100-hr subset from the training data per language. Later, we train streaming and non-streaming models. We report WERs on the full official test sets.

\subsection{Zipformer-Transducer Baseline}
\label{subsec:zipformer-transducer-exp}
We establish strong baselines by training non-streaming and streaming Zipformer transducer models~\cite{zipformer} from scratch, following the AMI recipe\footnote{\scriptsize\url{github.com/k2-fsa/icefall/tree/master/egs/ami}} from Icefall toolkit (we only use the IHM set).
The state-less Predictor~\cite{stateless-predictor} consists of an embedding layer followed by one 1-D CNN layer and joiner consists of 1 linear layer.
We use the default hyper-parameters from the recipes and train for 30 epochs.
See more training details in~\cite{zipformer}.
Beam search with width of 4 is used across all ASR decoding in this paper.

\subsection{XLSR-Transducer Training}
The XLSR-transducer model is constructed from the Icefall's Transducer recipe for AMI dataset adapted with the XLSR model from fairseq~\cite{ott2019fairseq}. The fine-tuning uses 
Scaled Adam~\cite{kingma2014adam} and a learning rate scheduler that consists of a 500-step warmup phase~\cite{vaswani2017attention} followed by a decay phase directed by number of steps and epochs. Furthermore, the model is optimized with pruned rnn-t loss~\cite{pruned-rnnt-loss,graves2012sequence}. The learning rate is set to $lr\!=1.25e^{-3}$ for AMI and $lr\!=5.0e^{-3}$ for CommonVoice. We train AMI and CommonVoice models for 10 and 20 epochs, respectively.

\begin{table}[t]
\caption{
WERs on the AMI eval set. On non-streaming decoding$^{\dagger}$ XLSR-Transducer yields significant WER reduction. On streaming decoding, the multi-chunk training (mutiple) provides significant gain w.r.t encoders trained from scratch with minimal degradation in non-streaming performance.
$^{\ddagger}$encoder-decoder model.
$^{\mathparagraph}$decoding chunk size 2000ms.
}
\centering
\label{table:ami-main}
\setlength\tabcolsep{4pt} 
\resizebox{1\linewidth}{!}{
\begin{tabular}{l c ccc}
    \toprule
    \textbf{Encoder} & \textbf{Chunk Size} & \multicolumn{3}{c}{\textbf{Chunk Size decoding}} \\
    \cmidrule(lr){3-5}
     & training & 320\,ms & 1280\,ms & full-att$^{\dagger}$ \\
    \midrule
    \multicolumn{4}{l}{\textbf{decoding: non-streaming ASR}} \\
    \cmidrule(lr){1-5}
    Whisper {\scriptsize large-v2 (1.6B)}$^{\ddagger}$~\cite{whisper} & - & - & - & 16.9 \\
    FastConformer {\scriptsize (1.1B)}~\cite{fastconformer}  & - & - & - & 15.6 \\
    Zipformer (70M)& - & - & - & 21.0 \\
    \midrule
    \multicolumn{4}{l}{\textbf{decoding: streaming ASR}} \\
    \cmidrule(lr){1-5}
    FastConformer {\scriptsize (114M)}$^{\mathparagraph}$~\cite{fastconformer-streaming}  & - & - & 24.2 & - \\
    Zipformer (70M) & multiple & 28.5 &  24.6 & 23.2 \\
    \midrule
    XLSR (300M) & full-att & 35.3 & 17.8 & \textbf{12.7} \\
    XLSR (300M) & 320\,ms & \textbf{17.1} & 15.0 & 14.2 \\    
    XLSR (300M) & 1280\,ms & 19.7 & 14.5 & 13.1 \\
    XLSR (300M) & multiple & 17.7 & \textbf{14.2} & \textbf{12.9} \\
    \bottomrule
    \end{tabular}
}
\vspace{-0.5cm}
\end{table}
\section{Results \& Discussion} 
\label{sec:results}

\begin{figure*}[t]
    \centering
    \begin{subfigure}[t]{0.3\textwidth}
        \includegraphics[width=1\textwidth]{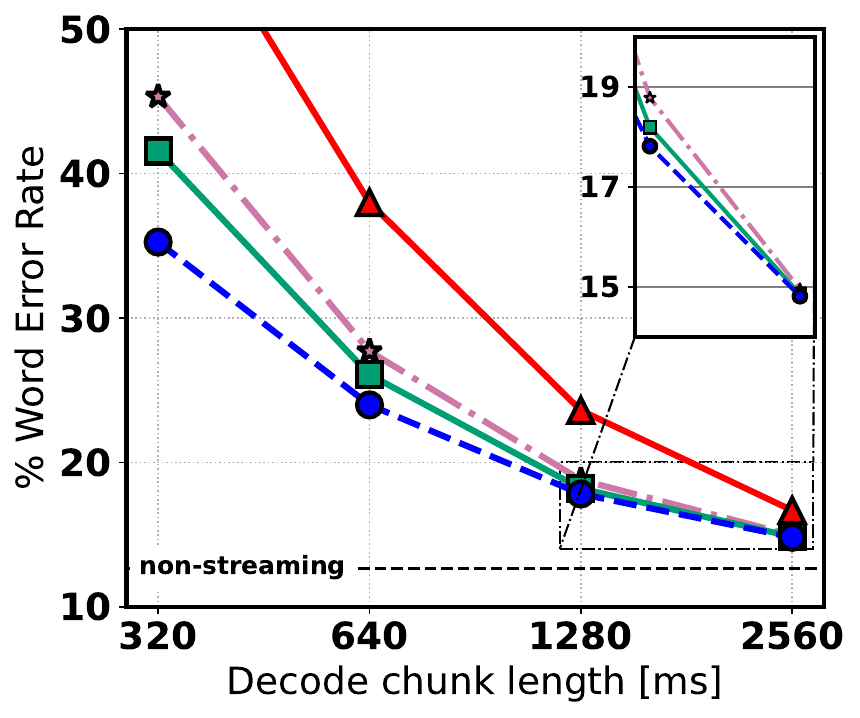}
        
        \caption{Non-streaming. }
        \label{fig:comparison-of-models}
    \end{subfigure}%
    \begin{subfigure}[t]{0.3\textwidth}
        \includegraphics[width=1\textwidth]{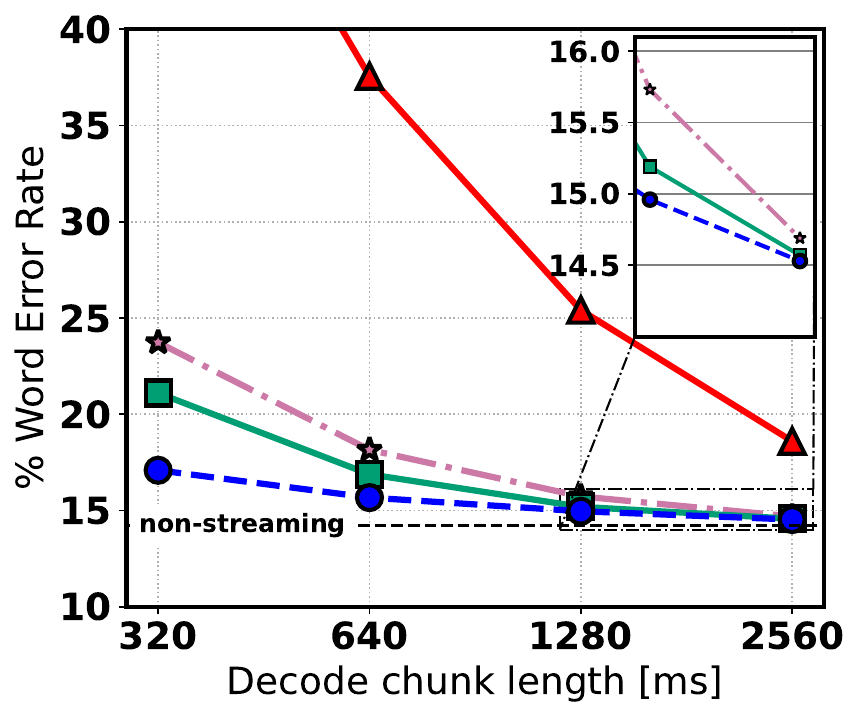}
        \caption{Chunk-size of 320ms and full left context.}
        \label{fig:comparison-of-heads}
    \end{subfigure}%
    \begin{subfigure}[t]{0.39\textwidth}
        \includegraphics[width=1\textwidth]{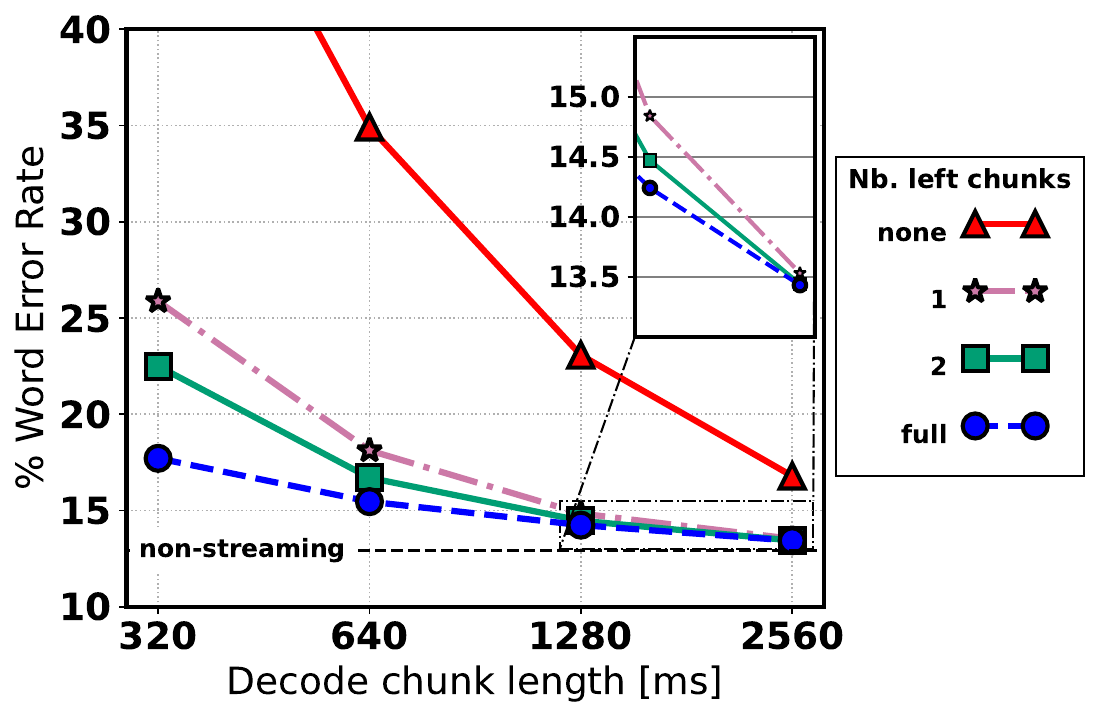}
        \caption{Multi-chunk streaming.}
        \label{fig:comparison-of-heads}
    \end{subfigure}%
    
    \caption{Plots of WERs on AMI eval set for XLSR-Transducer trained on three configurations (a, b and c) and decoded on multiple streaming scenarios. Note that adding one or more left-context chunks at decoding time reduces WERs dramatically.}
    \label{fig:WER-ami-streaming}
    \vspace{-0.31cm}
\end{figure*}

\noindent \textbf{Non-streaming ASR} \quad  We begin by benchmarking the XLSR-Transducer model for non-streaming ASR on the AMI dataset. Results are reported in the Table~\ref{table:ami-main} (\textit{full-attn}).
We compare against large open source foundational speech models.
It can be seen that the proposed XLSR-Transducer model achieves significant improvement in WERs.
Note that we do not finetune these open-source models on the AMI dataset because the primary goal of our work is to demonstrate best practices for utilizing SSL models in streaming ASR.
Next, we train a \textit{Zipformer} encoder based transducer model from scratch and observe that XLSR-Transducer yields 39\% relative improvement in WER.
It is clear that there are significant advantages of using pretrained encoders in TT setup for low resource ASR.

\noindent \textbf{Streaming ASR}  \quad First, we decode the non-streaming trained XLSR-Transducer model in streaming fashion by applying different masks (see \S\ref{subsec:streaming-xlsr-transducer}). Results are reported in the Table~\ref{table:ami-main}, where full left context is used during decoding.
The XLSR-Transducer achieves significant improvement over \textit{Zipformer} and \textit{FastConformer} based transducer models.
Despite the improvements, there is a significant degradation from non-streaming performance because the model was not trained for streaming.
When model was trained for streaming with full left context and decoded using chunk size of 320ms, the performance improves (35.3\% $\rightarrow$ 17.1\% WER) because of the train-test matched chunk size setting.
However, training with larger chunk sizes and decoding at 320ms degrades performance, showcasing the importance of context during self attention computation inside a chunk which the model may have learned during training.

Increasing the chunk duration during decoding (320ms $\rightarrow$ 1280ms) improves the performance monotonically \cite{rnnt-las-rescoring}, which is expected as larger context available for frame attention score computation.
Now, the streaming trained models are decoded in non-streaming which can serve as performance upper bound.
We observe that increasing chunk size during training improves the non-streaming performance and even when chunk size of 320ms is used during training, the results only degrade by 1.5\% in absolute WER.
When random chunk sizes are used during training, the gap is 0.2\% when compared with the best non-streaming ASR performance.
Overall, a streaming trained XLSR-Transducer model using random chunk sizes shows best WER when decoded in streaming fashion with 1280ms chunk duration and performance gap for non-streaming decoding is minimal.
Thus, a single model can be used for both streaming and non-streaming ASR.
\vspace{-0.3cm}
\begin{table}[t]
\centering
\caption{WERs of streaming XLSR-Transducer on five CommonVoice languages. Models are fine-tuned on random 100hr train subset and with multi-chunk training. \mbox{full-att:} non-streaming decoding. 
$^{\dagger}$CA is 28h long, while the rest 26h.
$^{\ddagger}$median (\@duration) in seconds.$^{\mathparagraph}$non-streaming training and decoding.}
\label{tab:CV-results}
\setlength\tabcolsep{3.5pt} 
\resizebox{0.99\linewidth}{!}{
\begin{tabular}{l cc | ccccc | c}
    \toprule
    
    \textbf{Lang$^{\dagger}$} & \multicolumn{2}{c|}{\textbf{Test Set}} & \multicolumn{5}{c|}{\textbf{Streaming model - chunk size [ms]}} & \textbf{full-att}$^{\mathparagraph}$ \\
    
    \cmidrule(lr){2-3}\cmidrule(lr){4-8}\cmidrule(lr){9-9}
    
    & \#utt & [50\%-dur]$^{\ddagger}$ & 320 & 640 & 1280 & 2560 & full-att & full-att \\
    \midrule
    CA & 16k & 6.1 & 17.5 & 15.2 & 13.9 & 12.9 & 12.0 & 10.7 \\
    BE & 15.8k & 5.7 & 20.0 & 17.5 & 15.9 & 14.8 & 13.8 & 13.7 \\
    ES & 15.5k & 6.1 & 17.7 & 15.0 & 13.5 & 12.2 & 11.3 & 10.8 \\
    FR & 16k & 5.7 & 24.3 & 21.6 & 20.0 & 18.7 & 17.6 & 17.1 \\
    IT & 15k & 6.3 & 18.5 & 15.9 & 14.3 & 13.1 & 12.1 & 11.5 \\
    \bottomrule

\end{tabular}
}
\vspace{-0.3cm}
\end{table}

\noindent \textbf{Streaming ASR with variable left context} \quad Using full left context during training and decoding will incur additional computation which may not always be desirable.
Limiting left context during training of streaming models resulted in significant degradation of results; therefore, we use full left context for all streaming XLSR-Transducer training.
Figure~\ref{fig:WER-ami-streaming} list WERs when the number of left context chunks are varied during decoding. 
Note that the left context duration is in multiple of chunk size.
Increasing the left context improves the performance for all training scenarios and chunk duration.
At the same time, a significant improvement is observed using just one chunk of left context.
Further increase in left context improves the performance but the reduction in WER per left context chunk is lower.
When the XLSR-Transducer is trained with multi-chunk streaming strategy and decoded with 1280ms chunk size, a relative improvement of only 4\% in WER is observed using full left context instead of one left context chunk.
Thus, a limited number of left context chunks should be enough for most real-world streaming ASR with XLSR-Transducer model.

\noindent \textbf{XLSR-Transducer on multiple languages} \quad We also train the proposed model on five non-English languages of CommonVoice~\cite{ardila2020common}. 
WERs are listed in Table~\ref{tab:CV-results} for multi-chunk streaming and full-attention non-streaming models. We see competitive WERs for models evaluated under different streaming conditions, with constant WERs improvement as chunk size increases; similar behavior is reported in~\cite{rnnt-las-rescoring}.
The upper-bound WERs are obtained with a model trained and evaluated in non-streaming fashion, i.e., last column of Table~\ref{tab:CV-results}. We note negligible WER degradation (up to 1.5\% absolute WER, worse CA; best BE) for full-attention decoding (\textit{full-att}) on streaming models vs. their non-streaming counterparts. This confirms the robustness of XLSR-Transducer on multiple languages.

\begin{table}[t]
\centering
\caption{WERs on AMI eval set for varied decoding settings, including attention sink. Adding attention sinks offer a better trade off than larger left context. \textcolor{blue}{(blue)} denotes relative WER reduction w.r.t no attention sink within same chunk and left context.
$^{\dagger}$Number of chunks.
$^{\ddagger}$Number attention sink frames.
}
\label{tab:attn-sink-exp}
\resizebox{0.99\linewidth}{!}{
\begin{tabular}{cc | ll}
    \toprule
    \multicolumn{2}{c|}{\textbf{Decoding settings}} & \multicolumn{2}{c}{\textbf{Decoding chunk-size}} \\
    \cmidrule(lr){1-2} \cmidrule(lr){3-4}
    
    \textbf{Left-context$^{\dagger}$} & \textbf{attn-sink$^{\ddagger}$} & \multicolumn{1}{c}{\textbf{320ms}} & \multicolumn{1}{c}{\textbf{640ms}} \\
    \midrule
    full & none & \multicolumn{1}{c}{17.7} & \multicolumn{1}{c}{15.5} \\
    \midrule
    \multirow{4}{*}{1} & none & 25.9 \quad {\scriptsize \textcolor{blue}{--}} & 18.1 \quad {\scriptsize \textcolor{blue}{--}} \\
    & 1 & 22.9~~{\scriptsize \textcolor{blue}{(+11.6)} } & 17.4~~{\scriptsize \textcolor{blue}{(+4.1)} } \\
    & 4 & 21.4~~{\scriptsize \textcolor{blue}{(+17.4)} } & 16.8~~{\scriptsize \textcolor{blue}{(+7.1)} } \\
    & 16 & 19.7~~{\scriptsize \textcolor{blue}{(+23.7)} } & 16.3~~{\scriptsize \textcolor{blue}{(+10.0)} } \\
    \midrule
    \multirow{4}{*}{2} & none & 22.5 \quad {\scriptsize \textcolor{blue}{--}} & 16.7 \quad {\scriptsize \textcolor{blue}{--}} \\
    & 1 & 20.9~~{\scriptsize \textcolor{blue}{(+7.1)} } & 16.4~~{\scriptsize \textcolor{blue}{(+1.8)} } \\
    & 4 & 20.0~~{\scriptsize \textcolor{blue}{(+11.1)} } & 16.2~~{\scriptsize \textcolor{blue}{(+3.0)} } \\
    & 16 & 18.9~~{\scriptsize \textcolor{blue}{(+16.0)} } & 15.9~~{\scriptsize \textcolor{blue}{(+4.8)} } \\
    \midrule
    4 & none & \multicolumn{1}{c}{19.8} & \multicolumn{1}{c}{15.9} \\
    \bottomrule
\end{tabular}
}
\vspace{-0.3cm}
\end{table}

\noindent \textbf{Improving streaming ASR with attention sinks} \quad
In theory, restricting left context chunks should lead to overall latency improvements for streaming ASR.
The recent observation of attention sinks phenomena \cite{attn-sink-main,attn-sinks-whisper}, where the transformer models learn to assign relatively higher attention scores to initial tokens, may help in reducing the overall computation required to decode one chunk of audio in streaming ASR.
WERs on the AMI dataset are reported in Table \ref{tab:attn-sink-exp} for the XLSR-Transducer model trained in multi-chunk setting.
For different chunk sizes and left contexts, we observe that increased frames for attention sinks improve the performance monotonically.
Specifically, for a smaller chunk of 320ms, using 1 left context chunk and 16 frames (i.e., 320ms) for attention sinks performs better than using 2 left context chunks by 12\% in relative terms despite attending over same number of frames.
We do not observe a significant reduction in WERs beyond a chunk size of 640ms.
Overall, our results show that it is better to use attention sinks than increasing left context chunks beyond 1 for improved performance.
We also run decoding with attention sinks on the CommonVoice dataset and observe similar trends but do not include a results table for brevity.

\section{Conclusions}
\label{sec:conclusions}
\vspace{-0.1cm}
In this paper, we demonstrate that using a self-supervised trained model as encoder in the transducer framework, termed as XLSR-Transducer, leads to significant improvement in WER on AMI dataset.
We explore various chunked masks and left context configurations to enable streaming decoding in XLSR.
Our findings across 2 datasets and 6 languages shows that the proposed model achieves streaming performance comparable to non-streaming ASR.

\section*{Acknowledgments}
This work was supported by the Idiap~\&~Uniphore collaboration project.

\bibliographystyle{IEEEtran}
\bibliography{references}

\end{document}